# An Integer Programming Formulation Applied to Optimum Allocation in Multivariate Stratified Sampling

**José André de Moura Brito**
National School of Statistical Sciences (ENCE/IBGE) - Brazil
e-mail: jambrito@gmail.com

**Gustavo Silva Semaan**
Federal University Fluminense – Instituto de Computação (UFF/IC) - Brazil
e-mail: gsemaan@ic.uff.br

**Pedro Luis do Nascimento Silva**
National School of Statistical Sciences (ENCE/IBGE) - Brazil
e-mail: pedronsilva@gmail.com

**Nelson Maculan**
Federal University of Rio de Janeiro (COPPE/UFRJ) - Brazil
e-mail: nelson.maculan@gmail.com

**Summary:** The problem of optimal allocation of samples in surveys using a stratified sampling plan was first discussed by Neyman in 1934. Since then, many researchers have studied the problem of the sample allocation in multivariate surveys and several methods have been proposed. Basically, these methods are divided into two class: The first involves forming a weighted average of the stratum variances and finding the optimal allocation for the average variance. The second class is associated with methods that require that an acceptable coefficient of variation for each of the variables on which the allocation is to be done. Particularly, this paper proposes a new optimization approach to the second problem. This approach is based on an integer programming formulation. Several experiments showed that the proposed approach is efficient way to solve this problem, considering a comparison of this approach with the other approach from the literature.

## 1. Introduction

Nowadays, much of the research applied by statistical institutes considers the adoption of a sampling plan. The sample survey allows obtaining estimates in relation to population parameters, based on a selected sample of this population (Lohr, 2010).

When you use a sampling plan, is necessary to balance the available budget for research and, the same time, to obtain a minimum level of precision for the estimates to be disclosed. An alternative to obtain these two requirements is the use of stratification techniques. Other words, with the meeting of members of the population in H homogeneous strata, it is possible to produce estimates with a higher level of accuracy, and this homogeneity measure based on the evaluation of an expression variance associated with a stratification variable previously chosen.

After defined the strata and a sample size n (defined in terms of the costs of both research and accuracy), independent samples are selected in each of these strata. In addition, there are some situations where the sample size is not defined a priori. In this case, when performing the allocation of samples (nh, h = 1, ...., M) to the strata being two purposes: (i) to Minimize a weighted sum of the variances associated with a set of m search variables; (2) to minimize the total sample size to be distributed among the strata. This way, the variation coefficients associated with these variables being equal to or lower coefficients of variation previously set (called cvs targets). In both cases, we have a multivariate allocation problem.

This paper presents a new methodology that meets the second purpose. This methodology is based on the application of integer programming formulation developed in R language. The paper is organized as follows: In section two we present some concepts of stratified sampling and a description of the Multivariate Allocation Problem. The section three brings the new methodology. The section four presents a small set of computational results, considering the application of the new methodology and the methodology proposed from the literature (Bethel (1989)).

## 2. Stratified Sampling and Optimal Allocation Problem

In Stratified Sampling (Cochran, 1977), a population U with N units is divided into M strata $E_1$, $E_2$, ..., $E_H$, formed by, respectively, $N_1$, $N_2$, ..., $N_H$ units. These strata do not overlap and together cover the entire population, in such a way that:

$$N_1 + N_2 + ... + N_h + ... + N_H = N \quad (1)$$

$$E_i \cap E_k = \emptyset, \ k = 1,...,H-1, i = k+1,...,H \quad (2)$$

$$\bigcup_{h=1}^{H} E_h = U \quad (3)$$

Once defined the strata, and a sample size n, are selected $n_h$ observations (independent samples) among $N_h$ observations available in each one of the strata $E_h$, where $n = n_1 + n_2 + ... + n_H$. In general, from this sample are identified information for a set of m search variables. Assuming that these variables are denoted by: $Y_1$, $Y_2$, ..., $Y_j$, ..., $Y_m$, the population variance in each of the strata for each of these variables is defined by:

$$S_{hj}^2 = \sum_{\forall i \in E_h}(y_{ij} - \overline{y}_{hj})^2, \ j = 1,...,m, h = 1,...,H \quad (4)$$

The $y_{ij}$ is the value of the i-th observation in stratum h associated with the j-th variable of research, and this variable is average at h-th stratum. Still about simple stratified sampling, the variance of the estimator of the total ($t_y$) (Cochran, 1977) for each of the m search variables is defined by:

$$V(t_{y_j}) = \sum_{h=1}^{H} \frac{N_h^2 \cdot S_{hj}^2}{n_h} \cdot (1 - \frac{n_h}{N_h}), \ j = 1,...,m \quad (5)$$

Once the values of $N_h$ and $Sh_j^2$ can be calculated from the definition of the strata, the amount of variance in equation (5) depends solely on the sample size to be allocated to nh stratum. This distribution is very important, because it is what will ensure the accuracy of the sampling procedure. But in practical terms, aiming to make an allocation, is necessary to balance the accuracy in relation to each of the research variables and the cost of research in relation to the sampling units to be investigated. According to the literature, there are two approaches that address this issue. The first considers the minimization of a weighted sum of the variances (or coefficients of variation) associated with the variables of research interest, set a sample size (n) maximum. The second should determine the sample size to be allocated to $n_h$ stratum, so that the whole sample is minimized, and the coefficient of variation estimates from the (variable $Y_j$) is less than or equal to a priori defined target cvs for these variables. We highlight the main works of literature that deal with two approaches: Kokan (1963), Kokan in Khan (1967), Huddleston, Claypool and Hocking (1970), Bethel (1989), Valliant and Gentle (1997), Khan and Ahsan (2003), Garcia and Cortez (2006), Kozak (2006), Day (2010), Khan, Ali and Ahmad (2011), Ismail and Nasser Ahmad (2011).

In this last case, this is equivalent to formulate the following mathematical programming problem, where $Y_j$ corresponding to the total of the j-th variable research, i.e.: $Y_j = \sum_{h=1}^{H} \sum_{i \in E_h} y_{ij}$

$$\text{Minimize} \sum_{h=1}^{H} n_h \quad (6)$$

$$\text{s.t.} \quad 1 \leq n_h \leq N_h, h = 1,...,H \quad (7)$$

$$\sqrt{V(t_{y_j})}/Y_j \leq cv_j \ j = 1,...,m \quad (8)$$

$$n_h \in Z_+ \ (h=1,...,H) \quad (9)$$

In this formulation, the objective function to be minimized (equation 6) is the sum of the sample sizes allocated to strata. The constraint given in equation (7) is allocated ensures that at least one sample unit to each of the strata and the number of allocated units will not exceed the size of the stratum. Already the restriction associated with equation (8) ensures that the ratio between the standard deviation of each variable and its respective search total is less than or equal to a target coefficient of variation setted in advance. Finally, the restriction of equation (9) ensures that the sample sizes allocated to strata are integers (restriction completeness of the problem). It adds further that the restrictions associated with equation (8) can be rewritten as follows:

$$\frac{V(t_{y_j})}{Y_j^2 .cv_j^2} \leq 1, \; j=1,...,m \quad (10)$$

Then, if continue developing the equation (5) and replacing it with the numerator of the constraints of type (8) yields:

$$\sum_{h=1}^{H} \frac{N_h^2 .S_{hj}^2}{n_h .Y_j^2 .cv_j^2} - \frac{N_h .S_{hj}^2}{Y_j^2 .cv_j^2} \leq 1, \; j=1,...,m \quad (11)$$

Since $N_h$, $S_{hj}$, $Y_j$ and $cv_j$ are obtained a priori, we can define the following constant: $p_{hj} = \frac{N_h^2 .S_{hj}^2}{Y_j^2 .cv_j^2}$

($h=1,...,H$, $j=1,...,m$). In this case, restrictions of type (11) take the following form:

$$\sum_{h=1}^{H} \frac{p_{hj}}{n_h} - p_{hj} \leq 1, \; j=1,...,m \quad (12)$$

A first alternative the resolution of the formulation defined by (6), (7), (9) and (12) would be the application of a method of non-linear programming (Bazaraa, Sheralli and Shetty, 2006; Luenberger and Ye, 2008) they worked with restrictions, such as the methods of penalties, multipliers, among others. Nevertheless, these methods produce sample sizes (solutions) which in general will not be integers. Furthermore, when performing rounding, there is no guarantee of global optimum (Wolsey, 1998). Alternatively, since the sample size should be integer (variables of the problem), one could think about the applying some integer programming method, for example, methods *Branch and Bound* (Land and Doig, 1960; Wolsey, 1998; Wolsey and Nemhauser, 1999). But the non-linearity of the constraints of type (12) with respect to variables $n_h$ (sample sizes) makes it impossible the application of these methods.

Given these observations, the following section provides a proposal for a new integer programming formulation which solve this problem and that is equivalent to the formulation defined by (6), (7), (9) and (12). More specifically, the resolution of this formulation is possible to produce the smallest sample size ($n_h$) which are allocated to whole strata and which satisfies the constraints (7) and (12). Other words, its resolution ensures the global optimum (Wolsey, 1998) with regard to the value of the objective function defined in (6).

## 3. The Proposed Formulation

Considering an optimization approach solve the problem defined by (6), (7), (9) and (12) involves determining which sample sizes $n_1, n_2,..., n_H$ be chosen from the sets defined by $A_h=\{1,2,3,...,N_h\}$ ($h=1,...,H$), in order to meet the constraints of the problem defined in the previous section and produce the minimum value for the objective function defined in (6). As optimization problem, there is a need to define the decision variables of the model. This sense, we introduce a binary variable $x_{hk}$ that takes the value "true" if the sample size $k \in A_h$ is allocated to stratum h (h=1,...,H). According the definition of this variable and the equations (6), (7), (9) and (12), we can write the following *Binary Integer Programming formulation* (BIP) (Wolsey and Nemhauser, 1999).

$$\text{Minimize} \sum_{h=1}^{H} \sum_{k=1}^{N_h} k.x_{hk} \tag{13}$$

$$\text{s.t.} \sum_{k=1}^{N_h} x_{hk} = 1, h = 1,...,H \tag{14}$$

$$\sum_{h=1}^{H} \sum_{k=1}^{N_h} \frac{1}{k}.x_{hk}.p_{hj} - p_{hj} \leq 1, \ j = 1,...,m \tag{15}$$

$$x_{hk} \in \{0,1\}, \ h = 1,...,H, k = 1,...,N_h \tag{16}$$

In this formulation, the restriction (14) ensures that, for each of the strata, there will be a variable $x_{hk}$ assuming exactly one value. This is equivalent to ensure the selection of only one k-value (sample size) for each set $A_h$ (h=1,...,H) and the constraint (15) is equivalent to the constraint (12) of the original formulation. This formulation does not address the issue of considering different costs in relation to the allocation of sample units to their respective strata, ie, the cost allocation are unitary. If there is such a need, the objective function given in (13) can be defined as follows:

$$\sum_{h=1}^{H} C_h \sum_{k=1}^{N_h} k.x_{hk} \tag{17}$$

$C_h$ corresponding to the cost and allocation of each sample to the stratum h (h=1,...,M).

Another issue that can be addressed with regard to the minimum sample size to be allocated to each of the strata, or $n_h \geq n_{\min}$ (h=1,...,H) ($n_{\min}=2,...$). This question can be considered from the inclusion of the following restriction:

$$\sum_{k=1}^{n_{\min}-1} x_{hk} = 0, h = 1,...,H \tag{18}$$

In order to illustrate this formulation, consider the following simple example (excluding (17) and (18)) which define three strata (H=3), $N_1=3$, $N_2=5$ and $N_3=4$ and only search variable (j=1). The proposed formulation would be as follows:

Minimize $1.x_{11} + 2.x_{12} + 3.x_{13} + 1.x_{21} + 2.x_{22} + 3.x_{23} + 4.x_{24} + 5.x_{25} + 1.x_{31} + 2.x_{32} + 3.x_{33} + 4.x_{34}$

Subject to:

$$x_{11} + x_{12} + x_{13} = 1 \quad (h=1)$$

$$x_{21} + x_{22} + x_{23} + x_{24} + x_{25} = 1 \quad (h=2)$$

$$x_{31} + x_{32} + x_{33} + x_{34} = 1 \quad (h=3)$$

$$p_{11}(1.x_{11} + \frac{1}{2}.x_{12} + \frac{1}{3}.x_{13}) - p_{11} + p_{21}(1.x_{21} + \frac{1}{2}.x_{22} + \frac{1}{3}.x_{23} + \frac{1}{4}.x_{24} + \frac{1}{5}.x_{25}) - p_{21} + p_{31}(1.x_{31} + \frac{1}{2}.x_{32} + \frac{1}{3}.x_{33} + \frac{1}{4}.x_{34}) - p_{31} \leq 1$$

$$x_{11}, x_{12}, x_{13}, x_{21}, x_{22}, x_{23}, x_{24}, x_{25}, x_{31}, x_{32}, x_{33}, x_{34} \in \{0,1\}$$

Generally, the *Integer Programming Formulation* (including PIBs) are solved by applying an implicit enumeration method as *Branch and Bound*. Methods like *Branch and Bound* (Wolsey and Nemhauser, 1999) find the optimal solution for *Integer Programming* efficiently, considering the resolution of a subset of problems associated with the feasible region of the problem. These methods were developed from the pioneering work of Land and Doig (1960).

## 4. Computational Results

This section provides a small set of computational results based on the application of the proposed formulation, and an enhanced version of the algorithm proposed by Bethel (Bethel, 1989). As regards to the formulation, we created a function (called BSM) on statistical software **R** (http://www.r-project.org) using the *lpSolve* package. The Algorithm of Bethel is available in the *SamplingStrata* package (also R). The computational experiments were performed on a computer with 24GB of RAM and processors of 3.40 GHz (I7). In order to evaluate the design, populations were used three different databases, which are: (1) POP_CAFE (Agricultural Census, 1998), (2) POP_FAZENDA_CANA e (3) POP_FAZENDA_GADO. The Table 1 provides some information on these populations: the strata of the population, in the research variables *Y* (m) and the total number of units (*N*).

Table 1 – Information about the used Databases.

| Population | H | m | N |
|---|---|---|---|
| POP_CAFE | 3 | 3 | 20472 |
| POP_FAZENDA_CANA | 4 | 3 | 338 |
| POP_FAZENDA_GADO | 7 | 2 | 430 |

The Tables 2 and 3 bring, respectively, the cvs and sample sizes (n) produced by applying the proposed formulation and the algorithm of Bethel. These tables shows that the new proposed formulation produced better solutions (sizes sample) compared to those produced by the algorithm of the literature.

Table 2 – Results of proposed formulation (BSM*).

| Population | $n_{BSM}$ | Cv Target | Coefficients produced by BSM | | |
|---|---|---|---|---|---|
| | | | j=1 | j=2 | j=3 |
| POP_CAFE | 2545 | 5% | 1.23 | 4.99 | 2.91 |
| POP_FAZENDA_CANA | 144 | 2% | 1.81 | 1.99 | 1.89 |
| POP_FAZENDA_GADO | 217 | 10% | 9.99 | 8.18 | |

*Proposed Formulation by Brito, Semaan e Maculan.

Table 3 – Results of the algorithm of Bethel.

| Population | $n_{Bethel}$ | Cv Target | Coefficients produced by algorithm of Bethel | | |
|---|---|---|---|---|---|
| | | | j=1 | j=2 | j=3 |
| POP_CAFE | 2546 | 5% | 1.23 | 5.00 | 2.91 |
| POP_FAZENDA_CANA | 146 | 2% | 1.78 | 1.96 | 1.84 |
| POP_FAZENDA_GADO | 219 | 10% | 9.89 | 8.04 | |

Certainly, it is necessary to evaluate a significant number of people, in order to quantify the gain from the application of the proposed formulation compared with the algorithm of Bethel and eventually with other algorithms from the literature. Thus, new computational experiments will be performed in a future work, considering populations of varying size (N) and with more research.